# Photon avalanche up-conversion in NdAl$_3$(BO$_3$)$_4$ nanoparticles excited at 1064 nm

Jefferson F. da Silva,[a] Rodrigo F. da Silva,[a] Emanuel P. Santos,[a] Lauro J. Q. Maia,[b] André L. Moura[a,c,*]

[a]Grupo de Física da Matéria Condensada, Núcleo de Ciências Exatas – NCEx, Campus Arapiraca, Universidade Federal de Alagoas, 57309-005, Arapiraca-AL, Brazil

[b]Instituto de Física, Universidade Federal de Goiás, Goiânia-GO, Brazil

[c]Programa de Pós-graduação em Física, Instituto de Física, Universidade Federal de Alagoas, Maceió-AL, Brazil

*Corresponding author. E-mail: andre.moura@fis.ufal.br

**Abstract**

We report efficient non-resonant ground state excitation at 1064 nm of trivalent neodymium (Nd$^{3+}$) ions in stoichiometric neodymium aluminum borate NdAl$_3$(BO$_3$)$_4$ nanoparticles, which are crystalline and, besides the large content of Nd$^{3+}$ ions, present excellent photoluminescence properties. Up-conversions (UCs) were observed and the energy pathways identified, as starting by multi-phonon assisted ground state absorption ($^4$I$_{9/2}$→$^4$F$_{3/2}$) and excited state absorption ($^4$I$_{11/2}$→$^4$F$_{3/2}$) with the population of the $^4$I$_{11/2}$ level by thermal coupling with the ground state. The excited state $^4$I$_{11/2}$ is also populated by relaxations of the Nd$^{3+}$ increasing the population of the $^4$F$_{3/2}$ level. Cross-relaxation among two Nd$^{3+}$ ions ($^4$F$_{3/2}$,$^4$I$_{9/2}$)→($^4$I$_{15/2}$,$^4$I$_{15/2}$) with subsequent phonon emission leads to two ions at the $^4$I$_{11/2}$ level every iteration triggering a photon avalanche mechanism which greatly enhances the efficiency of the UCs. Ladder thermal excitation $^4$F$_{3/2}$→[$^4$F$_{5/2}$,$^2$H$_{9/2}$]→[$^4$F$_{7/2}$,$^4$S$_{3/2}$]→$^4$F$_{9/2}$ was achieved, and the ground state relaxation from these levels provided emission at 880 nm, 810 nm, 750 nm, and 690 nm, respectively. Energy transfer UCs (Auger) between Nd$^{3+}$ ions at the $^4$F$_{3/2}$ level allowed population of the [$^2$G$_{3/2}$,$^4$G$_{7/2}$] from which relaxations to the $^4$I$_{9/2}$, $^4$I$_{11/2}$, and $^4$I$_{13/2}$ states provided emissions around 536 nm, 600 nm, and 660 nm, respectively. Associated to the nonradiative relaxations, we observed the heating of the nanoparticles (22 °C to 240 °C) with subsequent thermal enhancement of the frequency UCs due to the redistribution of population among coupled energy levels of the Nd$^{3+}$ ions. The present results have potential applications in super-resolution imaging and nanothermometry.

## 1. INTRODUCTION

Trivalent lanthanide ($Ln^{3+}$) ions possess the capability to absorb and emit radiation in a broad spectral range covering the ultra-violet, visible and infrared.[1] Because of that, the $Ln^{3+}$ ions are used for several applications due to their peculiar electronic structures.[2] One interesting mechanism to excite $Ln^{3+}$ ions is the photon avalanche (PA) which takes place when the excitation wavelength is not resonant with none ground state absorption (GSA) transition, but resonant with an excited state absorption (ESA) transition in a way that the GSA cross-section ($\sigma_{GSA}$) is much lower than the ESA absorption cross-section ($\sigma_{ESA}$), i.e., $\sigma_{GSA} << \sigma_{ESA}$.[3–5] A small population of the lower level of the ESA transition (intermediate level) promotes the ions to an upper energy level by ESA, and the excited ions can transfer the stored energy to neighbor ones by cross-relaxation promoting an increased number of ions to the intermediate level. Afterwards, these ions can absorb excitation photons each to be promoted to the more energetic one. Subsequent cross-relaxations can promote a huge number of ions to the intermediate state, and an avalanche mechanism can be established. In this way, the small absorption of the material to the excitation wavelength would increase by orders of magnitude and, consequently, the output luminescence due to the relaxations of ions in the more energetic level to lower lying ones.

Among the $Ln^{3+}$ ions, the photoluminescence properties of the neodymium ($Nd^{3+}$) ones have been deeply studied due to their rich energy level structure. For example, the $Nd^{3+}$ ions are feasible to be excited and emit radiation over the ultra-violet, visible and infrared spectral range, present several thermally coupled energy levels, and can exchange stored energy with neighbor ions of the same or different species. Excitation pathways of single, co- and triply-doped $Nd^{3+}$ ions materials involve up and down conversions, with temperature dependent absorption cross-section, cross-relaxations, excited state absorption, and thermal coupling of energy levels. PA in $Nd^{3+}$ based materials has been demonstrated in bulk materials ($YLiF_4$,[6,7] $LaCl_3$,[8,9] and $CeCl_3$[10]) at low temperature under visible excitation, and in nanoparticles ($LiLaNdP_4O_{12}$,[11] $NdVO_4$,[12,13] and $CeVO_4$[12]) at room temperature under near infrared excitation (~800 nm). Although the $Nd^{3+}$ ions have a strong radiative transition at around 1060 nm, these ions present very low GSA at this wavelength, but this wavelength is resonant with the ESA transition $^4I_{11/2} \rightarrow ^4F_{3/2}$ fulfilling one prerequisite to PA. To the best of our knowledge, there is only one report on PA in $Nd^{3+}$ doped materials under excitation around 1060 nm. The authors investigated PA in $NaYF_4$, $Y_2O_3$, $YGdO_3$, $YAlO_3$, $Y_3Al_5O_{12}$, $LiLaP_4O_{12}$, $Gd_2O_3$ and nanoparticles demonstrating an intense emission at around 880 nm ($Nd^{3+}$ transition $^4F_{3/2} \rightarrow ^4I_{9/2}$) with potential application in nanothermometry.[14]

A current trend involving Ln$^{3+}$ ions is the investigation of photoluminescence properties of nanoparticles with Nd$^{3+}$ ions synthetized by different chemical routes.[15–19] One interesting material is the Neodymium aluminum borate NdAl$_3$(BO$_3$)$_4$ which is obtained by replacing the Yttrium ions by the Nd$^{3+}$ in the YAl$_3$(BO$_3$)$_4$ crystalline structure.[20–24] Besides the large quantity of Nd$^{3+}$ ions which would favor quenching mechanisms, deleterious thermal effects, and crystalline structure modification, the NdAl$_3$(BO$_3$)$_4$ present excellent linear and nonlinear optical properties including relatively large quantum efficiency,[24] allows the laser (and random laser) operation[23,25–28] including with self-frequency conversions.[29,30]

Here we investigate an unconventional excitation of the Nd$^{3+}$ ions at 1064 nm using NdAl$_3$(BO$_3$)$_4$ nanoparticles. At this wavelength there is no resonant GSA, and the energy mismatch to the closest resonant ground state transition ($^4I_{9/2} \rightarrow {}^4F_{3/2}$) is large (≈1850 cm$^{-1}$). Also, the resonant ESA transition $^4I_{11/2} \rightarrow {}^4F_{3/2}$ is low probable at room temperature due to the large energy gap (≈1870 cm$^{-1}$) between the $^4I_{11/2}$ level and the ground state. Despite of that, we demonstrate that the excitation is feasible by phonon annihilation in order to compensate the energy mismatch. With the population of the $^4I_{11/2}$ level, the excitation photons are resonant with $^4I_{11/2} \rightarrow {}^4F_{3/2}$ ESA transition, succeeding a PA mechanism allowing an efficient population of the Nd$^{3+}$ ions to the $^4F_{3/2}$ level. The relaxation pathways of the Nd$^{3+}$ involved the multi-phonon generation increasing in the nanoparticles' temperature which provided the population redistribution of the Nd$^{3+}$ ions. The output photoluminescence consists of several wavelengths (visible and near infrared) whose energy pathways were identified.

## 2. EXPERIMENTAL

*a. Nanoparticles morphological and structural characterization*

The synthesis procedure of the NdAl$_3$(BO$_3$)$_4$ nanoparticles was performed by the polymeric precursor method as described in the Supplementary Material. The morphological and structural characterization was performed in Ref. [[29]]. It suffices to mentioning that the nanoparticles present crystalline structure of monoclinic cell with C2/c space group, as previously described and well-studied. The particle size distribution is broad and present peak at 55 nm. The largest phonon energy is 1360 cm$^{-1}$,[31,32] and the density of Nd$^{3+}$ ions is 5×10$^{21}$ ions/cm$^3$.

*b. Optical experiments*

The diffuse reflectance spectrum (DRS) of the nanoparticles powder was obtained using an UV/Vis/NIR spectrophotometer with BaSO$_4$ powders as reference. In the photoluminescence experiments, the excitation source was provided by a Nd:YAG laser operating at 1064 nm under continuous wave (cw) regime. The excitation power ($P_{exc}$) could be varied up to 2.0 W. The laser light was focused onto the powder which was gently pressed in a sample holder. The excited area at the powder surface was 0.06 mm$^2$. The generated light was collected by a pair of lenses, and focused onto a multimode optical fiber coupled to a spectrometer equipped with a charge-coupled device (ccd) allowing simultaneous spectral measurements from 330 nm to 1180 nm with 2.0 nm of resolution. The ccd integration time was fixed at 50 ms, allowing consecutive spectrum acquisition every 50 ms. A short-pass optical filter was used to reject the elastically scattered excitation laser light.

*c. Temperature measurements*

The interaction of the excitation laser with the nanoparticles involved non-radiative relaxations increasing the nanoparticles temperature. Simultaneously with the spectral measurements, the temperature was monitored with a thermal camera (detection range of -20 to 650 °C, accuracy of 1°C) allowing the association between spectrum and temperature of the nanoparticles.

**3. RESULTS AND DISCUSSIONS**

The absorbance of the NdAl$_3$(BO$_3$)$_4$ nanoparticles powder can be inferred by the diffuse reflectance spectrum [Fig. 1(a)] which presents several valleys corresponding to the ground state ($^4I_{9/2}$) absorption transitions of the Nd$^{3+}$ ions to the indicated energy levels. As represented by a vertical arrow, there is no resonant GSA for the excitation wavelength at 1064 nm. In spite of that, excitation of the nanoparticles at this wavelength provided temperature dependent up-converted near-infrared and visible emission bands centered around 880 nm, 810 nm, 750 nm, 690 nm, 660 nm, 600 nm, and 536 nm [Fig. 1(b) for an excitation power of 1.8 W] which we associate to the electronic transitions among the 4*f* states of the Nd$^{3+}$. A photograph of the nanoparticles powder at 240 °C is displayed as an inset in Fig. 1(b) which shows an intense yellowish color which could be seen by naked eyes even at room illumination. Even at the low spectral resolution of the acquisition system (2.0 nm), we observe sharp emission lines associated to electronic transitions within the measured bands, which correspond to transitions between Stark components of the mainfolds.

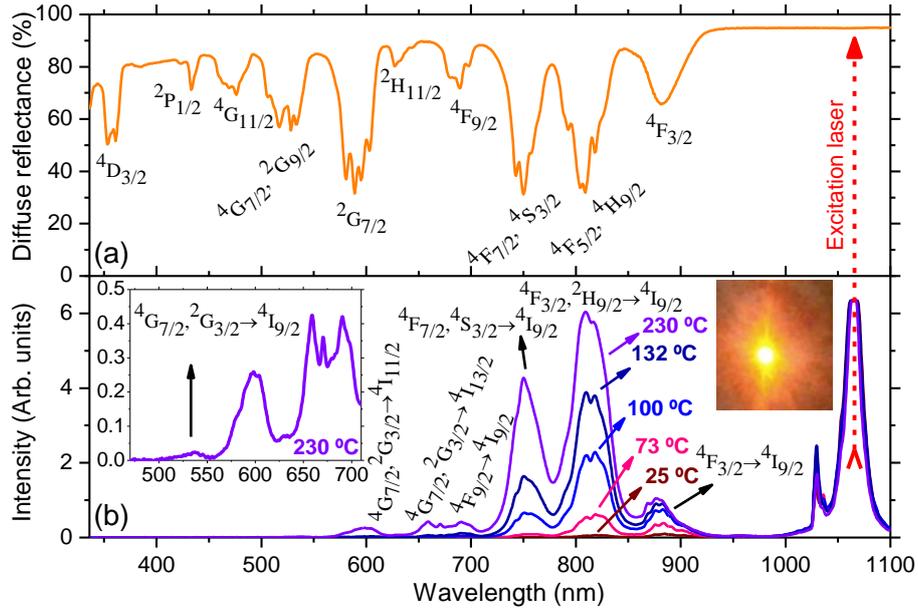

**FIG. 1.** NdAl$_3$(BO$_3$)$_4$ nanoparticles optical characterization. (a) Room temperature diffusive reflectance. The indicated states are the ending states of the ground state absorption transitions of the Nd$^{3+}$ ions. The up-arrow indicates the excitation wavelength ($\lambda_{exc}$ = 1064 nm) used in the photoluminescence experiments. (b) Photoluminescence spectrum under cw excitation at $\lambda_{exc}$ for an excitation power of 1.8 W for different nanoparticles temperature. The Nd$^{3+}$ ions transitions corresponding to each band are indicated. The inset on the right is a photograph of the powder at T = 240 °C.

With help of the available information for the NdAl$_3$(BO$_3$)$_4$ bulk crystal,[23] we represent the energy levels diagram of the NdAl$_3$(BO$_3$)$_4$ nanoparticles in Fig. 2 from where the energy pathways involved in the UCs in Fig. 1(b) are described. First, there is non-resonant absorption of excitation photons with the annihilation of phonons from the crystalline lattice to compensate the energy mismatch (≈1850 cm$^{-1}$) with the ground state transition $^4I_{9/2} \rightarrow ^4F_{3/2}$. Also, thermal coupling of the $^2I_{11/2}$ level with the ground state (ΔE ≈1870 cm$^{-1}$) allows the resonant ESA $^4I_{11/2} \rightarrow ^4F_{3/2}$. Due to the relatively large energy gap in both cases, the associated transition rates are very low at room temperature. In spite of that, once excited, the Nd$^{3+}$ ions at the $^4F_{3/2}$ can relax to lower lying levels, and the ground state relaxation transition $^4F_{3/2} \rightarrow ^4I_{9/2}$ provides the emission at around 880 nm. In addition, the Nd$^{3+}$ ions at the $^4F_{3/2}$ level can be excited to the [$^4F_{5/2}$,$^2H_{9/2}$] levels by phonon anihilation,[33–35] and by means of a ladder excitation to the [$^4F_{7/2}$,$^4S_{3/2}$] levels whose ground state relaxations provide radiation at 810 nm, and 750 nm, respectively. Similar thermal coupling under ESA at 1060 nm ($^4I_{11/2} \rightarrow ^4F_{3/2}$) as recently reported in LaPO$_4$:Nd$^{3+}$ nanoparticles.[36] Here we observed additional ladder thermal excitation populating the $^4F_{9/2}$ and upper lying levels. Thermal excitations between two given levels are represented by corrugated arrows in Fig. 2. The ground state relaxation $^4F_{9/2} \rightarrow ^4I_{9/2}$ corresponds to the emission at around 690 nm. On the other hand, the emissions at around 536 nm, 600 nm, and 660 nm are obtained by excitation of the Nd$^{3+}$ to the [$^2G_{3/2}$,$^4G_{7/2}$] levels which can occur by (1) ETU between two $Nd^{3+}$ at the $^4F_{3/2}$ levels with one of them relaxing to the $^4I_{13/2}$ level while the other being promoted to the $[^2G_{3/2},^4G_{7/2}]$ levels. Other possibility of populating the $[^2G_{3/2},^4G_{7/2}]$ levels is by (2) ESA of excitation radiation (ESAER) by $Nd^{3+}$ at the $^4F_{3/2}$ level. However, the ETU dominates over ESAER in highly doped crystals.[37] Based on previous works on $NdAl_3(BO_3)_4$ crystal,[24,38] the ETU is the dominating mechanism and the emissions at 536 nm, 600 nm, and 660 nm are assigned to the relaxations from the $[^2G_{3/2},^4G_{7/2}]$ levels to the $^4I_{9/2}$, $^4I_{11/2}$, and $^4I_{13/2}$ ones, respectively. Contribution from thermal coupling with lower lying levels to populate the $[^2G_{3/2},^4G_{7/2}]$ levels is not excluded. At large temperature and large ccd exposure time, a blue emission at around 480 nm was detected (Fig. S1), which we associated to thermal excitation $[^2G_{3/2},^4G_{7/2}] \rightarrow [^2G_{9/2},^4G_{11/2}]$ followed by the relaxation transition $[^2G_{9/2},^4G_{11/2}] \rightarrow ^4I_{9/2}$. At this point, it is worth mentioning that a previous study investigated $Ga_{10}Ge_{25}S_{65}$ glasses doped with $Nd^{3+}$ ions under pulsed (~7 ns) excitation at 1064 nm.[39] The observed emissions at around 535 nm, 600 nm, and 670 nm were attributed to the relaxation transitions $^4G_{7/2} \rightarrow ^4I_{9/2}$, $[^4G_{7/2} \rightarrow ^4I_{11/2}; ^4G_{5/2} \rightarrow ^4I_{9/2}]$, and $[^4G_{7/2} \rightarrow ^4I_{13/2}; ^4G_{5/2} \rightarrow ^4I_{11/2}]$, respectively, with two-photon absorption (TPA) from the $I_{9/2}$ to the $^4G_{7/2}$ and one photon absorption to the $^4F_{3/2}$ level followed by ETU among the $Nd^{3+}$ ions were identified as the excitation mechanisms. Here we also carried out experiments under pulsed regime, with the excitation laser operating at Q-switched and Mode-locked regimes, where the peak intensities are much larger than those in cw regime, and the obtained results were qualitatively equivalent to those obtained in cw regime (Fig. S2) ruling out TPA as a leading excitation mechanism in the $NdAl_3(BO_3)_4$ nanoparticles.

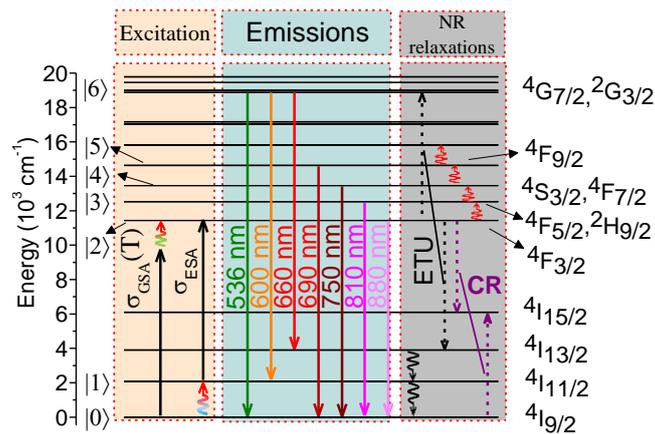

**FIG. 2.** Energy level diagram of the $NdAl_3(BO_3)_4$ nanoparticles. The vertical arrows indicate the possible energy pathways in the excitation of the nanoparticles. Thermal excitation between two given levels are represented by corrugated arrows. NR, CR, and ETU mean nonradiative relaxations, cross-relaxations and energy transfer up-conversion, respectively.

Generally, in photoluminescence experiments, the Input-Output power dependence ($P_{in}$-$P_{out}$) is given by $P_{out} = A(P_{in})^n$, with $A$ being a proportionality constant, and $n$ the number of excitation photons required for each absorption transition. According to the energy level diagram in Fig. 2, just one photon ($n$ = 1) is required to promote the $Nd^{3+}$ ions to the $^4F_{3/2}$ level, and for the emission at around 880 nm. However, $n$ varied with the $P_{exc}$ (Fig. 3) presenting a threshold power ($P_{th} \approx$ 1.1 W) from where the output intensity changed the behavior with a 3-fold enhancement. Based on Ref. [14], we associated it to the PA mechanism due to the resonant ESA $^4I_{11/2} \rightarrow ^4F_{3/2}$ transition triggered by a small initial population of the $^4I_{11/2}$ level by the thermal coupling with the ground state ($\Delta E \approx$ 1870 cm$^{-1}$) and several relaxation pathways of the $Nd^{3+}$ ions at the $^4F_{3/2}$ level. The radiative and multi-phonon relaxations of the $Nd^{3+}$ ions at the $^4F_{3/2}$ to lower lying levels ($^4I_{11/2}$, $^4I_{13/2}$, and $^4I_{15/2}$) increases the population of the $^4I_{11/2}$ level establishing an energy looping.[40] Additionally, cross-relaxations from the $Nd^{3+}$ ions at the $^4F_{3/2}$ to non-excited neighbor $Nd^{3+}$ ions (favored by the large quantity of $Nd^{3+}$ ions) **doubles** the number of ions at the $^4I_{15/2}$ which, by fast multi-phonon relaxations (favored by the large phonon energy), **doubles** the number of $Nd^{3+}$ ions at the $^4I_{11/2}$ level every iteration,[38] leading to strong population of the $^4F_{3/2}$ level by the resonant ESA. Notice that in this scenario, the PA condition $\sigma_{GSA} << \sigma_{ESA}$ is fulfilled. Corroborating with the proposed energy pathway, the other UC emissions demonstrated the threshold behavior (Fig. 3 shows the emission at 810 nm) due to the dependence of the corresponding emitting level with the population of the $^4F_{3/2}$ one.

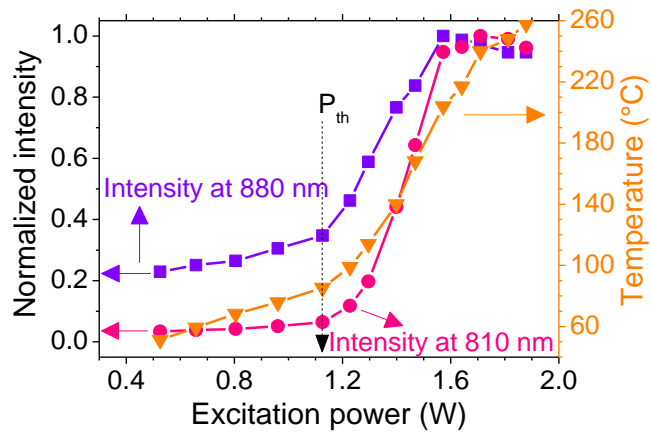

**FIG. 3.** Input-output power dependences for the up-converted emissions at 880 nm and 810 nm. A threshold excitation power of ≈1.1 W is well-defined, evidencing the photon avalanche mechanism. The excitation of the nanoparticles provided light-to-heat conversion, and the steady state temperature is also represented.

Associated to the nonradiative relaxations of the $Nd^{3+}$ ions, we observed the self-heating of the nanoparticles, i.e., temperature increasing induced by the excitation laser which provoked population redistribution among the $Nd^{3+}$ energy levels as described in the following.

The measurement procedure is detailed in the Supplementary Material. Figures 4(a)-(c) show the nanoparticles temperature, measured with the thermal camera, as a function of time of exposure to the excitation beam at $P_{exc}$ of 1.1 W, 1.5 W, and 1.8 W, respectively. The steady state temperature as a function of $P_{exc}$ is given in Fig. 3. Similar results were reported in NaYF$_4$:Nd$^{3+}$ nanoparticles under excitation at 795 nm, and the authors attributed the heating mainly to cross-relaxations among the Nd$^{3+}$.[41] Here it is known that large multi-phonon relaxation rates in NdAl$_3$(BO$_3$)$_4$ crystal are mainly due to the large phonon energy (1360 cm$^{-1}$),[24] which contributes to the light-to-heat conversion.

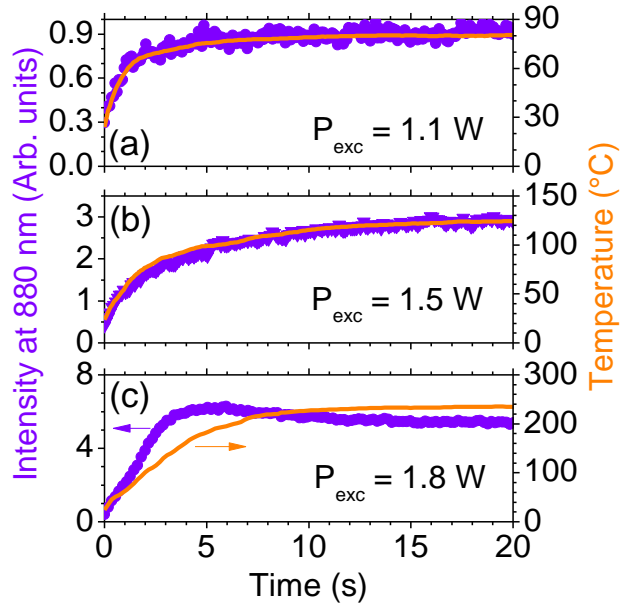

**FIG. 4.** Photoluminescence intensity at 880 nm and nanoparticles temperature as a function of exposure time to the excitation laser at the indicated power. The temperature increasing is associated to light-to-heat conversion due to non-radiative relaxations of the Nd$^{3+}$ ions.

The dynamics of the output luminescence at 880 nm and the nanoparticles temperature presented distinct behaviors depending on the $P_{exc}$ with relation to the PA threshold (Fig. 4). For $P_{exc}$ close to $P_{th}$, the time dependence of the temperature closely follows the output luminescence [Figs. 4(a) and (b)] which we associate to a looping mechanism where the nonradiative relaxations increases the nanoparticles temperature which increases the light absorption and promotes the population redistribution among the Nd$^{3+}$ ions energy levels. On the other hand, for $P_{exc}$ well-above $P_{th}$ [Fig. 4(c)] the build-up time of the photoluminescence at 880 nm is shorter than that of the temperature, and we associate this behavior to the PA dynamics which is faster at large excitation powers.[3,6,14]

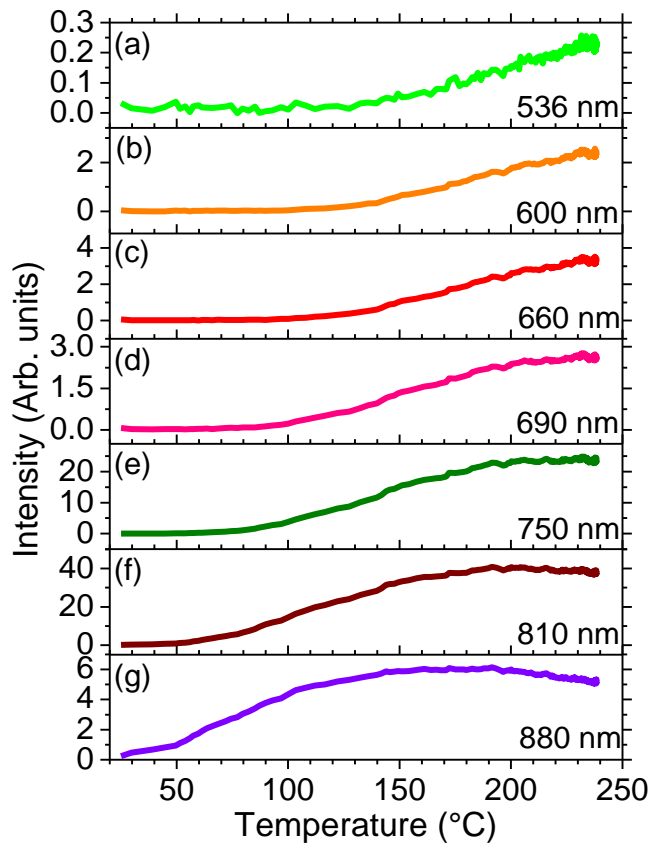

**FIG. 5.** Temperature dependence of the up-conversion emissions in NdAl$_3$(BO$_3$)$_4$ nanoparticles under excitation at 1064 nm for an average power of 1.8 W.

The change of the nanoparticles temperature due to the nonradiative relaxations of the Nd$^{3+}$ provoked pronounced modifications in the photoluminescence spectra, as shown in Fig. 1(b) for an excitation power of 1.8 W which is larger than the threshold for PA. The following features are observed in the output spectra. (1) large enhancements of the UC emissions. (2) changing in the shapes of the bands with a blue shift as the temperature increases. The peak intensity of the UC emission bands increased with the temperature presenting distinct dependences (Fig. 5). Enhancements of 440, 300, and 25 times are observed for the emissions around 750 nm, 810 nm, and 880 nm, respectively, with the temperature increasing from 22 °C to 240 °C. At room temperature (spectra obtained just after the interaction of the excitation beam with the nanoparticles), the other emissions were not intense enough to be detected by our measurement system, but were detected for temperature larger than 80 °C. One remarkable characteristic of the photoluminescence spectra is the spectral isolation of the emission bands, and also isolation from the excitation laser which is worth for application purposes. Notice that the temperature the emissions at 810 nm, 750 nm, and 690 nm presented distinct onset temperature for pronounced increasing corroborating the ladder thermal excitation mechanism, and the emissions at 536 nm, 600 nm, and 660 nm presented similar temperature dependences corroborating they are associated to transitions departing from the same energy level ([$^2G_{3/2}$,$^4G_{7/2}$]). A temperature dependent coupled rate equation model, which satisfactorily describes the experimental results, is given in the Supplementary Material.

## 4. SUMMARY AND CONCLUSIONS

In summary, we investigated an unconventional excitation of $Nd^{3+}$ ions at 1064 nm in $NdAl_3(BO_3)_4$ nanoparticles, which is out of resonance with ground state transitions. However, with the phonon annihilation, population of the $^4F_{3/2}$ is achieved with subsequent energy transfer to upper lying energy levels. It provided intense up-conversion (UC) emissions, which was enhanced by a photon avalanche (PA) mechanism with the population of the lower lying $^4I_{11/2}$ energy level, by which resonant excited state absorption increased the population of the $^4F_{3/2}$ level. Finally, the UC pathways involved several non-radiative relaxations increasing the temperature of the nanoparticles, which promoted population redistribution among the $Nd^{3+}$ ions energy levels. Promising applications of the PA is in super-resolution imaging,[42,43] and we foresee application of the UC emissions in nanothermometry due to the population redistribution among the $Nd^{3+}$ energy levels,[14,36,44,45] and also in hyperthermia by the light-to-heat conversion.[46]

## AUTHOR'S CONTRIBUTIONS

J.F.S. and A.L.M. conceived the project. J.F.S. performed the experimental measurements. All authors discussed the experimental results, implemented the rate equation model, and wrote the manuscript.

## SUPPLEMENTARY MATERIAL

See supplementary material for synthesis procedure of the $NdAl_3(BO)_4$ nanoparticles; response of the nanoparticles to a pulsed excitation beam (Q-switched and Mode-locked) at 1064 nm; blue emission band associated to the [$^2G_{9/2}$,$^4G_{11/2}$]→$^4I_{9/2}$ $Nd^{3+}$ ions transition; the measurement procedure of the photoluminescence simultaneously with the temperature; a temperature dependent coupled rate equation model which describes the experimental observations.

## ACKNOWLEDGEMENTS
We acknowledge financial support from the Brazilian Agencies: Fundação de Amparo à

Pesquisa do Estado de Alagoas (FAPEAL); Coordenação de Aperfeiçoamento de Pessoal de Nivel Superior (CAPES) – Finance Code 001, Fundação de Amparo à Pesquisa do Estado de Goiás (FAPEG), Financiadora de Estudos e Projetos (FINEP), Conselho Nacional de Desenvolvimento Científico e Tecnológico (CNPq) through the grant: Nr. 427606/2016-0; scholarships in Research Productivity 2 under the Nr. 303990/2019-8 and 305277/2017-0; National Institute of Photonics (INCT de Fotônica). E.P.S and R.F.S thank CNPq for their Scientific Initiation scholarship. We also thank to J. G. B. Cavalcante, M. A. Nascimento, and Professors C. Jacinto, M. V. D. Vermelho and A. S. Gouveia-Neto for technical support.